\documentclass[runningheads]{llncs}

\usepackage[utf8]{inputenc}
\usepackage{graphicx}
\usepackage{caption}
\usepackage{subcaption}
\usepackage{amsmath}
\usepackage{booktabs}
\usepackage{xspace} 
\usepackage{float}
\usepackage{marvosym}

\newcommand*{\etal}{\textit{et al}.\@\xspace}

\begin{document}

\title{Dynamics of scientific collaboration networks due to academic migrations}

\author{Pavlos Paraskevopoulos\inst{1}\orcidID{0000-0002-6858-7122} (\Letter) \and
Chiara Boldrini\inst{1}\orcidID{0000-0001-5080-8110} \and
Andrea Passarella\inst{1}\orcidID{0000-0002-1694-612X} \and
Marco Conti\inst{1}\orcidID{0000-0003-4097-4064}}

\authorrunning{Paraskevopoulos et al.}

\institute{CNR-IIT, Via G. Moruzzi 1, 56124, Pisa, Italy\\
\email{\{p.paraskevopoulos,c.boldrini,a.passarella,m.conti\}@iit.cnr.it}}


\maketitle

\begin{abstract}
    Academic migration is the change of host institution by a researcher, typically aimed at achieving a stronger research profile. Scientific features such as the number of collaborations, the productivity and its research impact tend to be directly affected by such movements. In this paper we analyse the dynamics of the collaboration network of researchers as they move from an institution to the next one. We specifically highlight cases where they increase and when they shrink, and quantify the dependency between the collaboration networks before and after such a movement. Finally, we drill down the analysis by dividing movements depending on the career stage of the researchers. The analysis shows a remarkable dynamism of collaboration network across migrations. Interestingly, not always movements result in larger collaboration networks, while the overall similarity between networks across movements is quite limited on average. Qualitatively, the same effects can be found at all career stages, while clearly the magnitude of them might vary. These results are based on a dataset extracted from Scopus, containing detailed scientific information for the publications of 84,141 researchers.
\end{abstract}

\section{Introduction}
\label{sec:intro}

Mobility is an important aspect of a researcher's life, affecting the career of the scientist in many ways~\cite{Fortunato2018,franzoni2014mover,petersen2018multiscale,sugimoto2017scientists,Verginer2018,Wang2019,Robinson-Garcia2019}. Through the change of host institutions, new career opportunities are chased, positions with higher prestige acquired, stronger collaborations can be created, and novel projects are started. As a result, the collaboration network, the productivity and the research impact of the studies of the researcher are possibly affected every time a researcher changes host institution.
Thus, mobility, both international and domestic, is rightly considered as a very important part of the research career. 
At first glance, the outcome of a movement may seem only beneficial. However, this is not always true. The downside to the creation of new opportunities is that old collaborations may collapse and old projects may be abandoned. The goal of this work is to assess the impact of academic migrations on the collaboration network of a researcher, also considering whether such features vary across different career stages. Moving towards this direction, we analyse how the size and composition of the collaboration network vary before and after an institution change. Finally, we investigate how the career duration affects the outcome of a movement in terms of collaboration network.

For our analysis, we use a dataset containing details for the scientific publications (extracted from the Scopus repository) of 84,141 mobile authors that have moved 561,389 times. The key element of our analysis is the author's ego network, which is defined as the set of co-authors (alters) of the given author (ego) while being at a certain institution.
For studying the effects of a host institution change on the ego network, we focus on the analysis of a set of characteristics such as the change of the ego network size, and the similarity between the ego networks before and after the movement. We first consider an overall exploratory analysis, where we study all movements together, and analyse (i) the percentage change of the ego network size, (ii) the overlap between the ego networks across each movement, (iii) how much of the old network ``survives" in the new one and (iv) to what extent the new network is built on old collaborations. We then drill down the analysis by separating movements that result in an increase of ego network size from those that result in a reduction. Finally, we separately analyse the same figures based on the different career stages of researchers.
The key results we highlight from our analysis are as follows:
\begin{itemize}
    \item On average, after a movement, the collaboration network tends to expand. However, the overlap between the old and new networks is not particularly high (Jaccard similarity between 10\%-20\%) and this result holds true both when the network size increases and when it decreases.
    \item On average, approximately 30\% of the old collaborations carry over to the new network, and they amount to $\sim 20\%$ of the latter. This turnover is related to the cognitive effort required to nurture collaborations: one cannot just add new relationships because time needs to be invested for their maintenance. In order to bring in new collaborations one needs to replace old ones.
    \item When the network size increases after a movement, still not all old collaborations are maintained. On average, only $36\%$ of old collaborations are carried over to the new network. When the network shrinks, this fraction is even smaller (around $14\%$) but the old collaborations relatively weight much more in the new network (around $36\%$).
    \item Movements tend, on average, to be more disruptive as the career progresses, partly because the networks after a movement are larger for more senior people but also because former collaborations tend to be curtailed more.
\end{itemize}

The rest of the document is organized as follows. In Section~\ref{sec:related} we present the related work. We present our dataset and its features in Section~\ref{dataset}. Section~\ref{sec:methodology} describes the metrics used for our analysis. We present our findings in Section~\ref{sec:results}, and draw conclusions in Section~\ref{sec:conclusion}.

\section{Related Work} 
\label{sec:related}
\vspace{-10pt}

Given the omnipresence of academic migrations in the careers of researchers, the movements of scientists have received a lot of attention in recent years. The migration flows from a country to another and how the language similarities affect them have been studied in~\cite{moed2013studying}. Their finding is that language similarity is one of the most important factors in relocation decisions.
Franzoni~\etal{} in~\cite{franzoni2014mover} focus on the effects of academic mobility on the impact factor of the published papers, finding positive effects on the career of researchers that migrate.
Movements to a higher- or a lower-ranked institution are considered in~\cite{deville2014career}. The main finding is that movement to a lower-ranked institution is affecting negatively the profile of a researcher, but the impact of authors that move to highly-ranked institutions remains the same. 
Sugimoto~\etal~\cite{sugimoto2017scientists} study the benefits of global researchers mobility. They found that mobile scholars always achieve higher citation rates than non-mobile ones. Mobile researchers do not cut their ties with their country of origin but instead work as bridges between different countries.
Mobility of researchers is also considered in~\cite{james2018prediction,urbinati2019hubs,vaccario2019mobility}, but from a different perspective, hence we do not discuss them further.

Closest to our work are the contributions by Petersen~\cite{petersen2018multiscale} and Arnaboldi~\etal~\cite{arnaboldi2016analysis}. Petersen~\cite{petersen2018multiscale} focuses on geographic displacements and their effect on the number of citations and collaboration network. His findings are threefold: (i) migration is associated with a significant churning in the collaboration network, (ii) the professional ties created after a migration event are less strong, (iii) career benefits of mobility are common to all ranks, not just to elite scientists. Despite considering the collaboration network, Petersen is mostly interested in the country to which the collaborators belong to, specifically focusing, e.g., on the effect of the country of previous collaborators on relocation decisions. Arnaboldi~\etal~\cite{arnaboldi2016analysis} also study the collaboration network of scientists, but they are mostly interested in uncovering the cognitive efforts behind collaboration maintenance. In addition, they do not consider the effects of mobility at all.

\vspace{-10pt}
\section{Dataset}
\label{dataset}
\vspace{-5pt}

The dataset that we use for our study is extracted from Scopus\footnote{The APIs used are the following: Author Search
API, the Author Retrieval API, and the Affiliation Retrieval API}, a popular academic knowledge repository managed by Elsevier, containing detailed records for the scientific publications of researchers from various disciplines. These records include the publications of a researcher, the authors of each publication and the affiliation of each of the authors at the time each study was published. Our dataset covers 84,141 authors. 
The average number of co-authors per paper in our dataset is 26, with a standard deviation of 219. 
The maximum number of co-authors in a paper is 5,563, while 96.8\% of the papers have less than 26 co-authors. 
Setting the latter as the threshold for assuming a meaningful collaboration, we dropped from the analysis the papers that have more than 26 authors. 

The affiliations specified in the publications by each author are used for the creation of affiliation time series, where each timeslot contains information related to the affiliation and the date.
The available timeslots of the time series are defined by the month and the year a paper was published and for every timeslot the author is considered to have only one available affiliation. In case a researcher has more than one affiliation declared in a timeslot's publications, we keep the one that appears more times in the timeslot. If we have affiliations with equal number of appearances, we keep the one that is declared as the author's affiliation in the closest in time (either in the past or the future) to the investigated timeslot.

Every affiliation change from an institution $inst_a$ to a different institution $inst_b$ for two consecutive timeslots defines a \textit{movement}\footnote{We are well aware that changes of affiliations as proxies for academic movements suffer from several limitations (e.g., authors with multiple affiliations could create spurious short detected movements). However, at the moment, it is the only approximation that allows researchers to study scientists' mobility at a large scale. We have in place some preprocessing aimed at mitigating the impact of such limitations (such as the removal of short movements, discussed later on). }. The duration of the \textit{institution stay} of a researcher at $inst_a$ is defined as the period between the first paper the researcher published under the affiliation $inst_a$ and the first paper published under the affiliation $inst_b$. As a result, we have \textit{institution stays} that have a minimum duration of 1 month. For the rest of the analysis, taking into consideration the example of the movement from $inst_a$ to $inst_b$, we define two networks for a given author. The first one is the old network of the ego ($net_{old}$), containing the co-authors of the ego during the stay at $inst_a$. The second one contains the co-authors of the ego during the stay at $inst_b$ ($net_{new}$). Finally, we define as \textit{research career duration}  the period between the first and the last publication of the author. 
Since different career stages may be affected differently by academic relocations, we extract the current career stage from the career duration using the following mapping, derived by considering the typical duration of the respective career stages: PhD students (PhD, 0 to 3 years), Young Researchers (YR, 3 to 6 years), Assistant Professors (AP, 6 to 10 years), Associate Professors (AsoP, 10 to 28 years), Full Professors (FP, 28 to 38 years) and Distinguished Professors (DP, 38+ years).
Our dataset consists of records that depict the activity of 84,141 authors that have at least one movement during their research career. Based on the career stage classification, our dataset consists of 26 PhD, 4,026 YR, 18,157 AP, 51,129 AsoP, 7,846 FP and 2,957 DP. The total number of movements that these authors have done is 561,389.

After analyzing the distribution of the \textit{institution stay} duration, we found that 343,312 movements have been done either before or after an institution stay of length less than 6 months (180 days). The occurrence of short institution stays could be caused either due to virtual migration of authors that have many affiliations or due to delays for the publication of accepted papers that contain the affiliation of the author at the time the paper was submitted. The remaining 218,077 movements (corresponding to stays above 6 months) have been done by 77,713 authors. These 77,713 authors are classified as: 13 PhD, 3,213 YR, 16,220 AP, 47,857 AsoP, 7,544 FP and 2,866 DP. In the following of the paper, most of the times we do not consider such short stays, hence the impact of, e.g., spurious short movements caused by multiple affiliations for the same author, is very limited. We explicitly mention when we consider also them, which is done primarily to assess that including or excluding such short movements does not change the essence of our results.


\vspace{-10pt}
\section{Methodology}
\label{sec:methodology}
\vspace{-5pt}

The movement of a researcher to a new institution may affect the collaborations of the researchers. While it might be intuitively assumed that movements result in a higher number of collaboration, this may not be always the case in reality. The main purpose of this paper is to analyse this aspect, and the resulting overlap between collaborations across movements. Specifically, we define a \emph{positive} movement as one that is associated with an increase of the ego network, while a \emph{negative} movement denotes one associated with a decrease.
Below, we define a set of metrics that capture the different ways a collaboration network might be affected by a movement. 
To this aim, we focus on a tagged movement~$j$ of researcher~$i$. 
Based on the notation introduced in the previous section, we denote with $net_{old}^{(ij)}$ the set of collaborators of author~$i$ before movement~$j$ and with $net_{new}^{(ij)}$ the set of collaborators of author~$i$ after movement~$j$. Since in the following we unambiguously refer to a tagged pair (author~$i$,~movement~$j$), we will drop the corresponding superscript.

First, we define a set of metrics focused on the size difference between the old and new collaboration network. We start with the ``difference of the network sizes" ($size_{diff}$), defined as the difference between the sizes of the new and the old ego networks.
We say that the effect of the movement is neutral when the $size_{diff}$ is equal to zero.

\begin{definition}\label{def:diff_size}
The ``difference of the network size" ($size_{diff}$) captures the difference between the size of the old and new network after a movement and is given by the following:
\begin{equation}
    size_{diff} = |net_{new}| - |net_{old}|.
\end{equation}
\end{definition}

\noindent
In order to further characterise this difference, we also calculate the ``difference ratio" ($diff_{ratio}$), which captures the relative change between the size of the old and new network. 

\begin{definition}\label{def:diff_ratio}
The ``difference ratio" ($diff_{ratio}$) measures the relative change between the size of the old and new network, and it is given by the following:
\begin{equation}
    diff_{ratio} = \frac{|net_{new}| - |net_{old}|}{|net_{old}|}.
\end{equation} 
\end{definition}

\noindent
Distinguishing between a positive movement (i.e., one that grows the ego network) and a negative one (i.e., one that shrinks), the third metric $size_{ratio}$ measures how much larger is the largest network with respect to the smaller one. 

\begin{definition}\label{def:size_ratio}
The ``Ratio of the network size`` ($size_{ratio}$) can be computed as follows:
\begin{equation}
    size_{ratio} = 
    \begin{cases} 
      \frac{|net_{new}|}{|net_{old}|} & | net_{new}| \geq |net_{old}| \\[1ex]
      - \frac{|net_{old}|}{|net_{new}|} & | net_{new}| < |net_{old}|
   \end{cases}
\end{equation}
\end{definition} 
Notice that, due to this definition, in the case of positive movements the $size_{ratio}$ is the multiplicative factor of increase of the new network size with respect to the old, while in the case of negative movements it is the division factor of decrease of the old network size with respect to the new.

While the previous group of metrics capture changes in network size, they do not shed light on how the composition of the network is modified after a movement. In order to assess the latter, we introduce the following additional metrics, that, taken together, fully characterise the compositional changes in the collaboration network. 


\begin{definition}\label{def:net_sim}
The similarity between the old and new collaboration network is measured with the Jaccard similarity, which can be obtained as follows:
\begin{equation}
    net_{sim} = \frac{\left| net_{old} \cap net_{new}\right|}{\left| net_{old} \cup net_{new} \right|}.
\end{equation}
\end{definition}


\begin{definition}\label{def:net_kept}
The ``Network kept" ($net_{kept}$) measures the fraction of old collaborators that keep collaborating with the tagged author after the movement, and it is given by the following:
\begin{equation}
    net_{kept} = \frac{\left| net_{old} \cap net_{new}\right|}{\left| net_{old} \right|}.
\end{equation}
\end{definition}

\begin{definition}\label{def:net_dep}
The ``Network dependency" ($net_{dep}$) measures the weight of old collaborations in the new network, and it is given by the following:
\begin{equation}
    net_{dep} = \frac{\left| net_{old} \cap net_{new}\right|}{\left| net_{new} \right|}.
\end{equation}
\end{definition}

\noindent
In short, $net_{sim}$ is a general measure of similarity between the two networks, while the other two capture the fraction of old collaboration that carry over to the new network and their relative weight in it. When needed, we will also denote with $|net_{inter}|$ the size of the intersection between the old and new network. 

\vspace{-10pt}
\section{Results}
\label{sec:results}
\vspace{-5pt}

In this section we investigate how academic movements affect the collaboration network of scientists. We present our general findings in Section~\ref{sec:results_general}. In Section~\ref{sec:results_pos_neg}, we focus our attention on the impact of positive or negative movements (i.e., movements that are associated with a growth or decrease in network size). Then, in Section~\ref{sec:results_career_stage} we study how the above results are affected by the career stage of researchers. Please note that the methodology used in the paper does not allow to establish clear causal relationships, but only co-occurrences, between academic movements and the observed changes in the collaboration network. The assessment of cause-effect patterns is left as future work.  

\vspace{-5pt}
\subsection{Sizes of networks and changes in collaborations across movements}
\label{sec:results_general}
\vspace{-5pt}

\begin{figure*}[t]
    \centering
    \includegraphics[width=\linewidth]{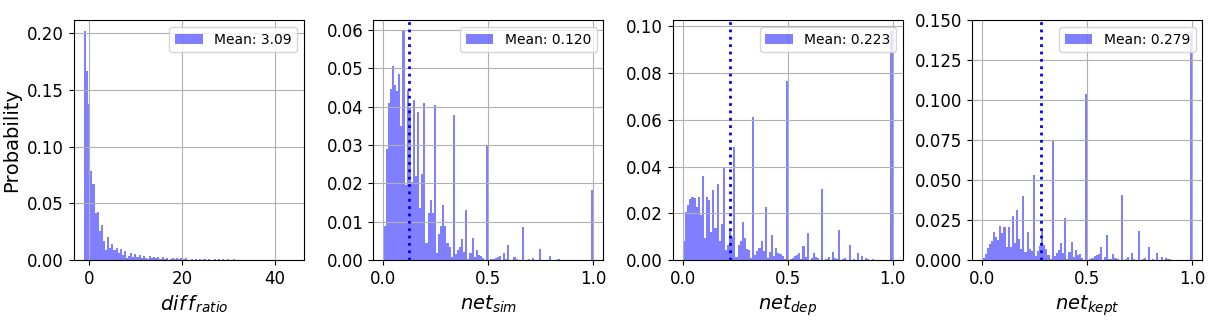}\vspace{-10pt}
    \caption{Distribution of the changes in size ($diff_{ratio}$) and composition of collaboration networks ($net_{sim}$, $net_{dep}$, $net_{kept}$) after a movement.}
    \label{fig:egoMods_All_a} \vspace{-10pt}
\end{figure*}

We start by studying what happens to the size of an author's collaboration network after they move from an institution to another one. For the moment, we do not filter out short inter-movement periods. Taking into consideration all the 561,389 movements in our dataset, we calculate the parameter $diff_{ratio}$ for every movement each author has recorded and we get the average $diff_{ratio}$ for each author. Averaging among all the authors, we find (Table~\ref{table:min_duration_comparison}) that the average $diff_{ratio}$ per author is 2.84, indicating that on average a movement is associated with an expansion of the network by 284\%.
This large average is due to some researchers whose network significantly grows after a movement (as also highlighted by the skewness in the $diff_{ratio}$ distribution in Figure~\ref{fig:egoMods_All_a}, which we will discuss later on). In Table~\ref{table:min_duration_comparison} we also report the median $diff_{ratio}$ value, which is around 8\%. Thus, as expected, movements are generally associated with a positive impact on the collaboration network. 

The fact that the number of collaborations increase after a movement tells us nothing about the composition of such network. To better understand this point, first we calculate the average intersection ($net_{inter}$) of the networks before and after a movement, for each author. From Table~\ref{table:min_duration_comparison}, we see that the average intersection of the authors is 1.93, indicating that (on average) a researcher keeps approximately 2 collaborations when moving to a new institution. In order to assess more precisely the similarity of the collaboration network before and after a movement, we compute the Jaccard similarity (Definition~\ref{def:net_sim}) at each movement, then we average these values across authors, obtaining an average Jaccard similarity of 0.11 (Table~\ref{table:min_duration_comparison}). Please note that a low similarity value was expected, since the Jaccard similarity also takes into account the difference in network size (and we have shown that the new network is approximately 4 times the old one, on average). However, if the old collaborations were fully retained, the similarity should be around $\frac{1}{4}$. The fact that it is lower implies that some old collaborations are removed. We will analyse this aspect more in detail later on in the section.

The second column of Table~\ref{table:min_duration_comparison} shows the values of the metrics described above but obtained discarding movements associated with short stays (smaller than 6 months). All the values change only slightly with respect to the first column. As short stays might also include non real migration events, and since including them or not does not make significant differences, in the following we consider the dataset where those short stays have been removed.

\begin{table}[t]
    \centering
    \setlength{\tabcolsep}{0.3em}
    \begin{tabular}{@{} l r r @{}}
    \toprule
                    & \textbf{No Threshold}               & \textbf{180 days}             \\ \midrule
    \textbf{\# of Movements}              & 561,389              & 218,077              \\ 
    \textbf{\# of Authors}                & 84,141               & 77,713               \\ \midrule
    & \multicolumn{2}{ c }{Difference Ratio ($diff_{ratio}$)}            \\ \cmidrule{2-3}
    \textbf{Avg}                          & 2.84                  & 3.09                 \\ 
    \textbf{Std}                          & 11.25                 & 10.4                 \\
    \textbf{Median}                       & 0.08                 & 0.36                 \\
    \midrule
    & \multicolumn{2}{c}{Jaccard similarity ($net_{sim}$)}            \\ \hline
    \textbf{Avg}                          & 0.11                 & 0.12                 \\ 
    \textbf{Std}                          & 0.16                 & 0.16                 \\ \midrule
    & \multicolumn{2}{ c }{Intersection ($net_{inter}$)}        \\ \hline
    \textbf{Avg}                          & 1.93                 & 2.15                 \\ 
    \textbf{Std}                          & 3                    & 3.3                  \\ \hline
    \end{tabular} \vspace{10pt}
    \caption{Comparison of old and new collaboration networks with and without short stays. The first column contains the values obtained from both short and long stays, the second column corresponds to the values obtained discarding inter-movement periods shorter than 6 months.} \vspace{-25pt}
    \label{table:min_duration_comparison}
\end{table}

\begin{figure*}[t]
    \centering
    \includegraphics[width=\linewidth]{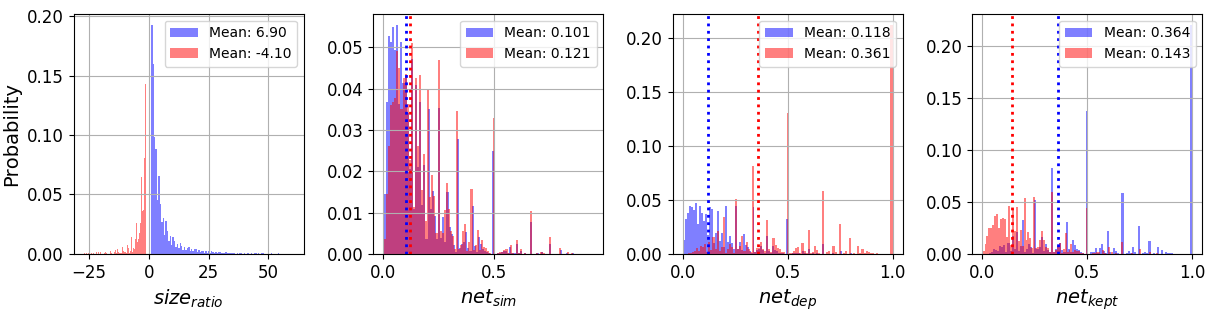} \vspace{-10pt}
    \caption{Distribution of the changes in size ($size_{ratio}$) and composition of collaboration networks ($net_{sim}$, $net_{dep}$, $net_{kept}$) after positive (blue) and negative (red) movements.}
    \label{fig:egoMods_All_b}\vspace{-10pt}
\end{figure*}




Figure~\ref{fig:egoMods_All_a} shows, in the first two charts, the distribution of the $diff_{ratio}$ metric and the Jaccard similarity $net_{sim}$, respectively. It is interesting to note that both distributions are quite spread: the size difference ratio could increase up to 20-40 times, while there are cases of complete overlap between the old and new network (corresponding to the peak at 1 for $net_{sim}$), even though the average overlap is in the order of $10\%$ only.
As discussed above, the Jaccard similarity values that we obtain hint at the fact that collaboration networks before and after a movement are not just different because of their size. In order to better investigate this aspect, we now focus on the analysis of $net_{kept}$ and $net_{dep}$, which capture specific variations in the composition of the collaboration network. The distributions of these metrics are shown in the third and fourth plot in Figure~\ref{fig:egoMods_All_a}, respectively.
We start with the analysis of the percentage of the old network that is retained after a movement (corresponding to the $net_{kept}$ metric). We find that on average, 27.9\% of the old network keeps collaborating with the ego also after the movement. 
The fact that more than 72\% of the old collaborations are dropped could probably be related to the cognitive effort required to nurture collaborations, restricting the ability to maintain the old ones.
On the other hand, looking at the $net_{dep}$ distribution, we observe that the collaborations overlapping between the old and new network are, on average $22.3\%$ of the new network. These two metrics allow us to better understand that the old collaborations can be seen as a good starting point after a movement, however, there is on average a very significant turnover associated with every movement. Interestingly, as we can see in Figure~\ref{fig:egoMods_All_a}, there are some extreme cases that keep active all of their old collaborations ($net_{kept}$ equal to 1), while some others form a new network that completely consists of old collaborations ($net_{kept}$ equal to 1). These are cases quite different from the ones described by the ``average" effect of movements.


\vspace{-10pt}
\subsection{Positive vs. negative movements}
\label{sec:results_pos_neg}
\vspace{-5pt}

The first results of our analysis indicate that, on average, after a movement the overall number of collaborations increases, even if 72\% of old collaborations are curtailed in the process. However, from the existence of negative values in the $diff_{ratio}$ (Figure~\ref{fig:egoMods_All_a}), it was clear that a movement would not always be associated with an increase in the number of collaborations a researcher has.
More precisely, only 121,980 (56\%) movements are associated with a positive effect on the network size, while 81,027 (37\%) movements are linked to its decrease (Table~\ref{table:summary_per_career}, column ``All''). The remaining 7\% of the movements results in equal size networks for both institution stays before and after a movement.

Figure~\ref{fig:egoMods_All_b} and Table~\ref{table:net_stats_per_career} (column ``All") show the analysis of positive and negative movements separately. Specifically, in Figure~\ref{fig:egoMods_All_b} blue histograms correspond to positive movements, while red histograms correspond to negative movements.
Analyzing, in Table~\ref{table:net_stats_per_career} column ``All'', the ratio of the network size ($size_{ratio}$) for the positive and the negative movements (Definition~\ref{def:size_ratio}), we found that the positive movements correspond to new collaboration networks that are on average 6.9 times the size of the old network. On the other hand, the negative movements would result into networks of a size 4.1 times smaller than the network size prior the movement. 
As far as it concerns the similarity $net_{sim}$, it was similar for both positive and negative movements with values around 0.1. Interestingly, 40,412 positive movements and 27,190 negative movements have \textit{0} Jaccard similarity, indicating that after these movements no prior collaboration remained active.

The \emph{dependency} of the new network on the old network (i.e., the ratio between the intersection and the size of the new network -- Definition~\ref{def:net_dep}), after a positive movement, is 11,8\%, while it jumps to the 36,1\% for the negative movements (Table~\ref{table:net_stats_per_career}). Remember that this metric captures how much the new network is based on old collaborations. While this increase is  expected (since in the case of negative movements the new network size is smaller), it is, however, interesting to note that also in this case not all old relationships are kept. As it is clear from Figure~\ref{fig:egoMods_All_b}, the percentage of times where this is the case (i.e., the $net_{dep}$ metric is equal to 1) is only about $20\%$. 

The $net_{kept}$ metric allows us to analyse the dual phenomenon, i.e., what is the proportion of old collaborations that are kept in the new network. In case of positive movements, 36.4\% of old collaborations are kept, while this metric for the negative movements is 14.3\% (Table~\ref{table:net_stats_per_career}). The fact that this metric is higher in case of positive movements could also be expected, as the new network is bigger, and therefore there is ``more space" to accommodate also old collaborations. It is quite interesting to note, however, that despite such additional ``space" only about $20\%$ of authors keep all their old collaborations when moving (depicted by a value of $net_{kept}$ equal to 1 in Figure~\ref{fig:egoMods_All_b}).

\subsection{Impact of migrations at the different career stages}
\label{sec:results_career_stage}

\begin{table}[t]
    \hspace*{-0.5cm}
    \scriptsize
    \centering
    \begin{tabular}{@{} l l l l l l l @{}}
    \toprule
    \textbf{Sample Size}   & \multicolumn{6}{c}{218,077 Movements || \textbf{Pos}: 121,980 / \textbf{Neg}: 81,027 / \textbf{Neutral:} 15,070}    \\ \toprule
                  & \textbf{All}            & \textbf{YR}             & \textbf{AP}             & \textbf{AsoP}        & \textbf{FP}           & \textbf{DP}           \\ \midrule
    \textbf{\# Authors}       & 77,713         &  3,213 (4.1\%)    & 16,220 (20.9\%)         & 47,857 (61.6\%)       & 7,544 (9.7\%)       & 2,866 (3.7\%)   \\ 
    \textbf{\# Movements}     & 218,077        & 4,360 (2\%)    &  29,093 (13.3\%)       &    136,105 (62.4\%)    &  32,392 (14.9\%)     & 16,113 (7.4\%)  \\ 
    \textbf{\# Pos Mov}       & 121,980 (56\%)     & 2,111 (48\%)     & 15,696 (54\%)     & 77,923 (57\%)  & 17,754 (55\%)     & 8,491 (53\%)   \\ 
    \textbf{\# Neg Mov}       & 81,027 (37\%)    & 1,765 (40\%)    & 11,326 (39\%)   & 49,359 (36\%)   & 12298 (38\%)  & 6,275 (39\%)  \\ \bottomrule
    \end{tabular} \vspace{5pt}
    \caption{General dataset statistics per career group.}
    \label{table:summary_per_career}\vspace{-30pt}
\end{table}

In this section, we take into account how the career stage of a researcher affects the results discussed so far. Indeed, it is reasonable to expect that an academic movement has a different effect on the network of collaborations if you are a PhD student or a full professor. It is not immediately clear in which direction this effect applies, though. On the one hand, for junior scientists, establishing new collaborations might be more difficult. On the other hand, the incentives for them to establish a widespread professional network are definitely stronger.  
Taking this into consideration, we split our dataset into the 6 career groups defined in Section~\ref{dataset} (we drop the PhD students from the analysis because they are too few). As shown in Table~\ref{table:summary_per_career}, the bulk of the authors in our dataset can be classified as associated professors, and they also account for the majority of movements. 
%
The proportion between positive and negative movements is very similar, regardless of the career stage, with the positive movements being around 55\% and the negative movements being between 36\% and 40\%. The only career group that slightly differs is the group of Young Researcher, who have a 48\% of movements resulting into increased network size. This slightly smaller percentage could be due to the short career duration of the authors that results into short institution stays. The latter, reinforces our idea that short institution stays are not ideal for forming an adequate and representative network.

    
    
    
    

\begin{table}[t]
    \footnotesize
    \centering
    \setlength{\tabcolsep}{1em}
    \begin{tabular}{@{} r r r r r r r @{}}
    \toprule
                  & \textbf{All}            & \textbf{YR}             & \textbf{AP}             & \textbf{AsoP}        & \textbf{FP}           & \textbf{DP}           \\ \toprule
    & \multicolumn{6}{c}{Difference Ratio ($diff_{ratio}$)}           \\ \cmidrule{2-7}
    \textbf{All}           & 3.09           &   0.99         &  1.67          &  3.15        &  4.13       &  3.65  \\ \midrule

    & \multicolumn{6}{c}{Ratio of the network size ($size_{ratio}$)}           \\ \cmidrule{2-7}
    \textbf{Pos}           & 6.9            & 3.5           & 4.5             & 6.7         & 9            & 8.4         \\ 
    \textbf{Neg}           & -4.1           & -3.2          & -3.35           & -4          & -4.8         & -5.3        \\ \midrule
    \multicolumn{7}{c}{Jaccard similarity ($net_{sim}$)}      \\ \cmidrule{2-7}
    \textbf{All}           & 0.12           & 0.18           & 0.14           & 0.12        & 0.11         & 0.1          \\ 
    \textbf{Pos}           & 0.1            & 0.14           & 0.12           & 0.1         & 0.09        & 0.08         \\ 
    \textbf{Neg}           & 0.12           & 0.16           & 0.13           & 0.12        & 0.11         & 0.1          \\ \midrule
    & \multicolumn{6}{c}{Network Kept ($net_{kept}$)} \\ \cmidrule{2-7}
    \textbf{All}           & 0.27           & 0.32           & 0.3            & 0.28        & 0.27         & 0.25         \\ 
    \textbf{Pos}           & 0.36           & 0.4            & 0.37           & 0.37        & 0.36         & 0.34         \\ 
    \textbf{Neg}           & 0.14           & 0.19           & 0.16           & 0.14        & 0.13         & 0.12         \\ \midrule
    & \multicolumn{6}{c}{Network Dependency ($net_{dep}$)} \\ \cmidrule{2-7}
    \textbf{All}           & 0.22           & 0.3            & 0.25           & 0.22        & 0.22         & 0.21         \\ 
    \textbf{Pos}           & 0.12           & 0.17           & 0.14           & 0.11        & 0.1          & 0.1          \\ 
    \textbf{Neg}           & 0.36           & 0.41           & 0.37           & 0.36        & 0.36         & 0.35         \\ \bottomrule
    \end{tabular} \vspace{5pt}
    \caption{Average network statistics per career group.}
    \label{table:net_stats_per_career}
    \vspace{-30pt}
\end{table}

From the network size standpoint, we see in Table~\ref{table:net_stats_per_career} that a movement would (on average) be associated with increase in the network size of the author, regardless of the career group the authors belongs to. This increase is around 100\% for YR, 167\% for AP, 315\% for AsoP, 413\% for FP, and 365\% for DP.  Given the skewness of the $diff_{ratio}$ distribution, the median values are smaller but they substantially confirm this growth trend. Thus, it seems that the longer the career duration, the higher the expansion of the network.
The fact the DP grow their networks slightly less than FP may be due to the fact that at the career stage of distinguished professor expanding collaborations might be less relevant. If we distinguish between positive and negative movements, the $size_{ratio}$ metric tells us that, for positive movements, the trend described above is substantially confirmed: the collaboration network increasingly grows from YR to FP. Then, it faces a minor setback for DP, which are, as discussed above, a corner case. Quite interestingly, the trend in the negative movements is the same in terms of absolute values of $size_{ratio}$ but the implications are opposite: when authors curtail collaborations after a movement, they do so, much more when they are senior than junior. 

Let us now focus on the variations of the collaboration network composition, at the different career stages. The variation of the average Jaccard similarity values across the career stages is less pronounced than the variation in network size. The overall low values are substantially expected based on the considerations in Sec.~\ref{sec:results_general} about the network size difference. It is also interesting to note that the difference between the case of positive and negative movements is barely noticeable. 
The $net_{kept}$ values in Table~\ref{table:net_stats_per_career} tell us that the different career stages do not impact significantly on the fraction of the old network that is retained after a movement: indeed, the average $net_{kept}$ is always around 0.3. Slightly more variation is observed for the average $net_{dep}$, corresponding to the fraction of old collaborations in the new network, but the difference is still limited. We now consider the combined effect of $net_{kept}$ and $net_{dep}$ for positive and negative movements. We substantially confirm the findings discussed in Section~\ref{sec:results_pos_neg}: when authors grow their network after a movement, they retain an important fraction of the old network, which, obviously, does not weight as much in the new network (which is larger). When the collaboration network shrinks, there is anyhow a significant collaborations turnover, with $\sim15\%$ of old relationships retained, which now account for $\sim 35\%$ of collaborations in the new network.  

\vspace{-15pt}
\section{Conclusions}
\label{sec:conclusion}
\vspace{-8pt}

The changing of host institutions is a fundamental part of the research career, yet the impact that such movements have on a scientist's network of collaborations is not completely understood. In this study, we have presented a preliminary investigation of the potential effects a movement has on the network of a researcher from the standpoint of network size and composition of the collaboration network. For our study, we have used records of scientific publications published by 84,141 researchers, extracted from the well-known platform Scopus.
Our findings highlight that on average, the collaboration network after a movement tends to be larger. Yet, the overlap between the old and new networks is not particularly high (Jaccard similarity between 10\%-20\%), regardless of the movement being positive or negative. In terms of how much the old network ``survives'' in the new one, we found that, on average, approximately 30\% of the old collaborations carry over to the new network. Old collaborations amount typically to $\sim 20\%$ of the new network. Distinguishing between positive and negative movements, when the network size increases after a movement, still, not all old collaborations are maintained. On average, only $36\%$ of old collaborations are carried over to the new network. When the network shrinks, this fraction is even smaller (around $14\%$) but the old collaborations relatively weight much more in the new network (around $36\%$). The career stage of scientists does not affect significantly the above results. Movements tend, on average, to be slightly more disruptive as the career progresses, partly because the networks after a movement are larger for more senior people but also because former collaborations tend to be curtailed more.


\subsubsection*{Acknowledgements.}

This work was partially funded by the SoBigData++, HumaneAI-Net, MARVEL, and OK-INSAID projects. The SoBigData++ project has received funding from the European Union's Horizon 2020 research and innovation programme under grant agreement No 871042. The HumaneAI-Net project has received funding from the European Union's Horizon 2020 research and innovation programme under grant agreement No 952026. The MARVEL project has received funding from the European Union's Horizon 2020 research and innovation programme under grant agreement No 957337. The OK-INSAID project has received funding from the Italian PON-MISE program under grant agreement ARS01 00917.

The work of Pavlos Paraskevopoulos was supported by the ERCIM Alain Bensoussan Fellowship Program.

\bibliographystyle{splncs04}
\bibliography{refs}

\begin{thebibliography}{10}
\providecommand{\url}[1]{\texttt{#1}}
\providecommand{\urlprefix}{URL }
\providecommand{\doi}[1]{https://doi.org/#1}

\bibitem{arnaboldi2016analysis}
Arnaboldi, V., Dunbar, R.I., Passarella, A., Conti, M.: Analysis of
  co-authorship ego networks. In: International Conference and School on
  Network Science. pp. 82--96. Springer (2016)

\bibitem{deville2014career}
Deville, P., Wang, D., Sinatra, R., Song, C., Blondel, V.D., Barab{\'a}si,
  A.L.: Career on the move: Geography, stratification, and scientific impact.
  Scientific reports  \textbf{4}, ~4770 (2014)

\bibitem{Fortunato2018}
Fortunato, S., Bergstrom, C.T., B{\"{o}}rner, K., Evans, J.A., Helbing, D.,
  Milojevi{\'{c}}, S., Petersen, A.M., Radicchi, F., Sinatra, R., Uzzi, B.,
  Vespignani, A., Waltman, L., Wang, D., Barab{\'{a}}si, A.L.: {Science of
  science}. Science (New York, N.Y.)  \textbf{359}(6379),  eaao0185 (mar 2018).
  \doi{10.1126/science.aao0185}

\bibitem{franzoni2014mover}
Franzoni, C., Scellato, G., Stephan, P.: The mover's advantage: The superior
  performance of migrant scientists. Economics Letters  \textbf{122}(1),
  89--93 (2014)

\bibitem{james2018prediction}
James, C., Pappalardo, L., S{\^\i}rbu, A., Simini, F.: Prediction of next
  career moves from scientific profiles. arXiv preprint arXiv:1802.04830
  (2018)

\bibitem{moed2013studying}
Moed, H.F., Aisati, M., Plume, A.: Studying scientific migration in scopus.
  Scientometrics  \textbf{94}(3),  929--942 (2013)

\bibitem{petersen2018multiscale}
Petersen, A.M.: Multiscale impact of researcher mobility. Journal of The Royal
  Society Interface  \textbf{15}(146),  20180580 (2018)

\bibitem{Robinson-Garcia2019}
Robinson-Garcia, N., Sugimoto, C.R., Murray, D., Yegros-Yegros, A.,
  Larivi{\`{e}}re, V., Costas, R.: {The many faces of mobility: Using
  bibliometric data to measure the movement of scientists}. Journal of
  Informetrics  \textbf{13}(1),  50--63 (feb 2019).
  \doi{10.1016/J.JOI.2018.11.002},
  \url{https://www.sciencedirect.com/science/article/pii/S1751157718300865}

\bibitem{sugimoto2017scientists}
Sugimoto, C.R., Robinson-Garc{\'\i}a, N., Murray, D.S., Yegros-Yegros, A.,
  Costas, R., Larivi{\`e}re, V.: Scientists have most impact when they're free
  to move. Nature News  \textbf{550}(7674), ~29 (2017)

\bibitem{urbinati2019hubs}
Urbinati, A., Galimberti, E., Ruffo, G.: Hubs and authorities of the scientific
  migration network. arXiv preprint arXiv:1907.07175  (2019)

\bibitem{vaccario2019mobility}
Vaccario, G., Verginer, L., Schweitzer, F.: The mobility network of scientists:
  Analyzing temporal correlations in scientific careers. arXiv preprint
  arXiv:1905.06142  (2019)

\bibitem{Verginer2018}
Verginer, L., Riccaboni, M.: {Brain-Circulation Network: The Global Mobility of
  the Life Scientists}. arXiv (October) (2018),
  \url{http://eprints.imtlucca.it/4072/1/EIC{\_}WP{\_}4{\_}2018.pdf}

\bibitem{Wang2019}
Wang, J., Hooi, R., Li, A.X., Chou, M.h.: {Collaboration patterns of mobile
  academics: The impact of international mobility}. Science and Public Policy
  \textbf{46}(3),  450--462 (jun 2019). \doi{10.1093/scipol/scy073},
  \url{https://academic.oup.com/spp/article/46/3/450/5281269}

\end{thebibliography}


\end{document}